 \documentclass[final,1p,times]{article}

\usepackage{amssymb,graphicx}

\begin{document}



\title{Simulation of space acquisition process of pedestrians using Proxemic Floor Field Model}


\author{
Takahiro Ezaki
\thanks{
             Department of Aeronautics and Astronautics, School of
	     Engineering, The University of Tokyo.
             e-mail:ezaki@jamology.rcast.u-tokyo.ac.jp}
, Daichi Yanagisawa
\thanks{
	     Research Center for Advanced Science and Technology,
The University of Tokyo (RCAST), and Japan Society for the Promotion of Science (JSPS).
}
, Kazumichi Ohtsuka
\thanks{
	Research Center for Advanced Science and Technology, The University of Tokyo, and Economics and Social Research Institute (ESRI), Cabinet office, Government of Japan.
}
\\and Katsuhiro Nishinari
\thanks{
	Research Center for Advanced Science and Technology, The University of Tokyo (RCAST), Department of Aeronautics and Astronautics, School of Engineering, The University of Tokyo and 
        PRESTO, Japan Science and Technology Agency.
}
}
\date{}

\maketitle

\begin{abstract}

We propose the Proxemic Floor Field Model as an extension of the Floor Field Model, which is one of the successful models describing the pedestrian dynamics. 
Proxemic Floor Field is the Floor Field which corresponds to the effect of repulsion force between others.
By introducing the Proxemic Floor Field and threshold, we investigate the process that pedestrians enter a certain area. 
 The results of simulations are evaluated
by simple approximate analyses and newly introduced indices. The difference in pedestrian behavior due to the disposition of the entrance is also confirmed, namely, 
 the entrance in the corner of the area leads to the long entrance time because of the obstruction by pedestrians settling on the boundary cells.
\end{abstract}


\section{Introduction}
Study on pedestrian dynamics has attracted many physicists over the past few years\cite{rev1,rev2,nagatani}.
We can see many collective phenomena induced from the complexity of the system of pedestrian flow.
In order to describe these phenomena, some models have been suggested and have achieved success 
in some situations. Helbing \itshape et al. \upshape designed the \itshape social force model \upshape \cite{SocialForce,social2,social3}, which represents pedestrian movement 
using differential equations. Interaction between pedestrians is demonstrated by introducing ``social force" - which acts between others. Using this model, we can reproduce phenomena such as arching
 at an exit, lane formation in counter flow, and oscillations of flow directions at narrow doors where pedestrians are crowding in both sides.
 
 \par Cellular Automaton (CA) based models  are also often used for their simplicity. 
In CA based models such as the \itshape lattice gas model \upshape \cite{lattice} and \itshape floor field model \upshape \cite{floor},
 time and space are discretized and simple update rules are introduced. These characteristics of CA based models are suited for numerical simulations.

\par The floor field model is one of the major CA based models that describe the pedestrian dynamics stochastically on latticed cells. In Ref.\cite{floor}, two types of \itshape Floor Field (FF) \upshape  which control hopping probability are introduced.
One is the \itshape Static Floor Field (SFF) \upshape which controls the ``destination of pedestrians". The other is the \itshape Dynamic Floor Field (DFF) \upshape
 which describes the characteristic that pedestrians tend to follow their predecessor. 
 In addition, other studies introduced the effects of wall \cite{extendedFF}, force \cite{extendedFF2} and collisions \cite{extendedFF3} among pedestrians etc.,
 and many valuable results have been obtained. 
 
\par Although CA based models have been studied extensively for describing the evacuation process \cite{floor,evacuation,evacuation2} and counter flow \cite{lattice,lattice2,floor}, etc., few studies have been done
 for the \itshape proxemic \upshape problem. \itshape Proxemics \upshape is the term introduced by Hall \cite{proxemics} referring to 
 the study of set measurable distances between people as they interact. Hall proposed four types of distance range: Intimate distance, Personal distance,
  Social distance, and Public distance. Hall also summarized the effects of proxemics as \itshape``Like gravity, the influence of two bodies on each other is inversely
   proportional not only to the square of their distance but possibly even the cube of the distance between them." \upshape 
 This repulsive force has been introduced into a few CA based models\cite{was}. 

 In this paper, the characteristics of the proxemic behavior are integrated into the floor field model.
 We just introduced the \itshape Proxemic Floor Field (PFF)\upshape, which describes the 
 proxemic behavior of pedestrians, and the \itshape threshold\upshape, which describes 
the ``motivation to move" dependent on the crowdedness of the neighborhood of each pedestrian.
 By using this model, we investigate \itshape inflow process \upshape, in which the proxemic effect is predominant. Inflow process is the situation where pedestrians 
 enter a certain area and acquire their own space; this can be observed on elevators, trains, attractions in theme parks, and many other places.
 This process is universally seen in the situations where pedestrians temporarily stay in a limited area. 
 The study of this process can be applied to the design of buildings, infrastructures, and so on.
 In such situations, after entering the area, pedestrians look for the location where they can keep their own space and are not intruded upon by other pedestrians.
 This movement is driven by the repulsion force between others. After they successfully acquire the location, they stops walking.
 We also have to consider the attractiveness of locations. For example, the space in front of the exit in the train carriage is valuable for passengers in a hurry who do not want to
 be obstructed during boarding, passengers who carry bulky items, and so on. On the other hand, the place may be uncomfortable for 
 passengers not in a hurry because many passengers intrude their space every time the train arrives at the stations.

 Though there are some CA based models which consider the ``force" of pedestrians\cite{extendedFF2,Gipps,Comp,force2}, they mainly focus on the local effect related to 
 physical contact and thus they are not suited for simulating the non-local repulsion dominant in the inflow process. Moreover, 
 in the \itshape social distance model\upshape \cite{was} dealing with the long range effect of social force, proxemic force does not affect walking pedestrians, 
 so this model cannot simulate inflow process either. Thus, the Proxemic Floor Field model is a new model capable of simulating inflow process. 
 Furthermore, it should be noted that, although they seem to be just an opposite processes, the inflow dynamics are more complicated 
 than the evacuation ones because the destination is not clear in the inflow process; therefore we have to set the driving force on each pedestrian. 
 This is why the inflow process has been less studied, compared to evacuation process.
  Of course, this model can also be used for other situations, which adds the effect of proxemic force to previous CA based models. 
 
 We evaluated simulation results of the time required for all the pedestrians to enter the room by mean field analyses and confirmed the difference due to
 the disposition of the entrance.   
 We focus on not only the inflow process, but also the \itshape steady state\upshape ; the state in which every pedestrian has stopped. 
 We evaluated the steady state by introducing a couple of new indices to represent the disposition of the pedestrians.

\section{Model}
\label{Model}
In this section we summarize the update rules of Proxemic Floor Field Model.
 The space is discretized into cells which can either be empty or occupied by one pedestrian. Each pedestrian
can move to one of the unoccupied next-neighbor cells (or stay on the present cell) at each discrete time step
$t\longrightarrow t+1$ depending on Floor Fields and threshold as explained below.

In the simulation of the inflow process, the Static Floor Field $S$ represents the attractiveness / unattractiveness of each cell for pedestrians.
We can control the pedestrian's behavior such as ``stay on a comfortable place (local maximum / minimum of $S$)" ,
``approach the destination (gradient of $S$)", or ``never stay on an uncomfortable place (local minimum / maximum of $S$)", 
by setting the appropriate Static Floor Field. It should be noted that this Static Floor Field is often set as the value dependent on
the distance from the exit in the context of the evacuation process. 
The Dynamic Floor Field $D$ represents the characteristic that pedestrians tend to follow their predecessors. This tendency is 
observed in crowded or emergency situations. This is implemented by virtual pheromone, which has its own dynamics of diffusion and decay 
controlled by the parameters $\alpha$ and $\delta$. Each pedestrian drops a boson on the present cell as he/she walks, and this boson decays 
or diffuses with probability $\alpha$ and $\delta$ respectively in each time step. The Dynamic Floor Field $D$ is the number
 of bosons remaining on the cell at each time step. These two types of Floor Fields have been historically used as fundamental and essential components of the Floor Field Model.

As explained below, transition probabilities of Floor Field Model are modified by following components newly introduced in this paper.  
\subsection{Proxemic Floor Field}
\begin{figure}[htbp]
 \begin{minipage}[t]{0.32\hsize}
  \begin{center}
  $r_{ij}^1$\\
   \includegraphics[width=3cm,clip]{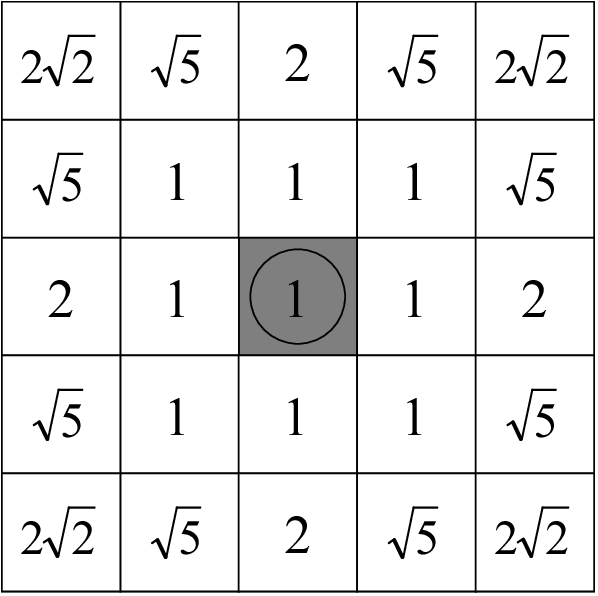}
  \caption{Distances from the 1st pedestrian in the center cell. Distances between two neighboring cells and within the cell itself are defined to be 1.}
  \label{r}
  \end{center}
 \end{minipage}
 \hfill
 \begin{minipage}[t]{0.32\hsize}
 \begin{center}
 $P^1_{ij}$\\
  \includegraphics[width=3cm,height=3cm,clip]{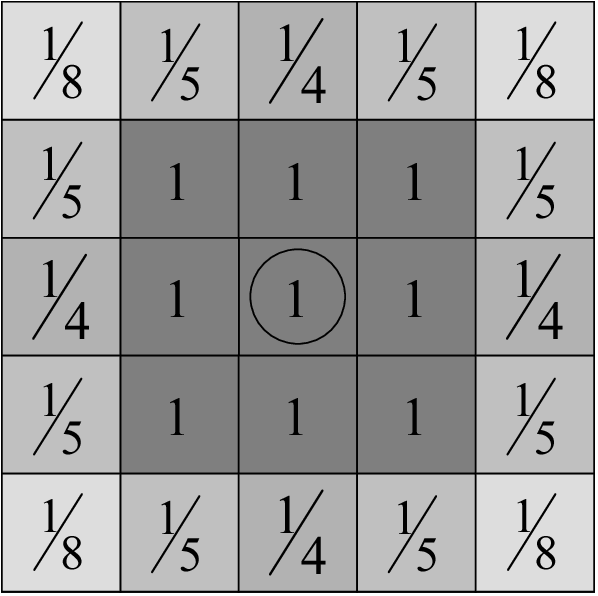}
  \caption{Proxemic Floor Field generated by one pedestrian.}
  \label{fig:three}
 \end{center}
 \end{minipage}
 \hfill
 \begin{minipage}[t]{0.32\hsize}
 \begin{center}
 $\sum_{k=1}^2{P^k_{ij}}$\\
  \includegraphics[width=3cm,height=3cm,clip]{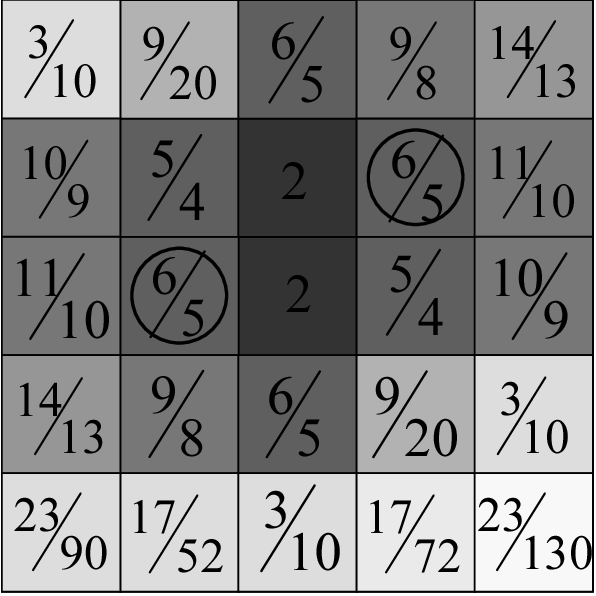}
  \caption{Proxemic Floor Field generated by two pedestrians.}
  \label{fig:two}
 \end{center}

 \end{minipage}
\end{figure}

Proxemic Floor Field represents the effect of repulsion which acts between others according to their distance. 
As mentioned in Ref.\cite{proxemics}, repulsion for small distance is much greater than that for large distance. 
In this paper we define the Proxemic Floor Field $P_{ij}^k$ generated by each pedestrian $k$ as (\ref{pro1}), and $P_{ij}$ is gained by taking sum of 
each $P_{ij}^k$ for all the pedestrians. Here, $r_{ij}^k$ is the distance between $k$-th pedestrian and cell-$(i,j)$. This distance is defined to be $1$ for
 the neighboring cells and its own cell, while other cells are measured in the $L2$ norm (see Figure \ref{r}), for the naturalness of the model. 
By adopting this distance, Proxemic Floor Field generated by a pedestrian is uniformly assigned to neighboring cells, and this PFF does not influence its transition probability. (See figs. \ref{fig:three} and \ref{fig:two}.)

\begin{eqnarray}
\label{pro1}  P_{ij}^k &=& \frac{1}{\left(r_{ij}^k\right)^2}\\ 
\label{pro2}  P_{ij} &=& \sum_k{P_{ij}^k}
\end{eqnarray}

\subsection{Threshold}
Pedestrians stop walking after they acquire a suitable position in trains, elevators, etc.
Here, we introduce a threshold which is related to the decision making about walk or stop: when the threshold is larger than the gradient of Floor Fields, 
pedestrians tend to stay on their present cells. 
Threshold $\Theta$ is given by (\ref{theta}) for each pedestrian. The subscript $O$ corresponds to its present cell and 
 $W_{ij}$ denotes \itshape total \upshape Floor Field , i.e., the weighted sum of Floor Fields. 
The magnitude of $W_{ij}$ controls the pedestrian movement (explained by transition probabilities in the next subsection). 
The greater the $W_{ij}$, the lower threshold. This corresponds to the fact that
pedestrians in crowded area tend to move to more attractive places (low threshold), and 
that pedestrians in spacious areas tend to ignore the small gradient of \itshape total Floor Field \upshape (high threshold).

\begin{eqnarray}
\Theta &=& \Theta_{max}\exp{\{-k_TW_O\}}\label{theta}\\
W_{ij} &=& -k_DD_{ij} + k_SS_{ij} + k_PP_{ij}
\end{eqnarray}
where,\\
\begin{tabular}{rp{29em}}
$\Theta_{max} \in [0,\infty)$&:the coefficient which controls the magnitude of threshold.\\
$k_T \in (0,\infty)$&:the sensitivity parameter which controls the steepness of threshold against $W_{ij}$.\\
$k_S \in [0,\infty)$&:the sensitivity parameter which weights the effect of Static Floor Field \\
$k_D \in [0,\infty)$&:the sensitivity parameter which weights the effect of Dynamic Floor Field \\
$k_P \in [0,\infty)$&:the sensitivity parameter which weights the effect of Proxemic Floor Field, which is newly introduced in this paper explained below. \\
\end{tabular}

\subsection{Basic update rules}
Pedestrians follow the rules below in the Proxemic Floor Field Model.
Note that each $S_{ij}$ is given and $D_{ij}=0$ for all cells at time step $t=0$ 
\begin{enumerate}
\item $D_{ij}$ is modified by the rules of diffusion and decay.
\item Each pedestrian chooses one target cell based on the transition probabilities $p_{ij}$ determined by (\ref{transition}) in the order numbered (sequential update).
\begin{eqnarray}\label{transition}
p_{ij}&=&\left\{ 
\begin{array}{ll}
N\exp{(-W_{ij} - \Theta)} & : transition\;  to\; the\; neighboring\; cells \\
N\exp{(-W_{O})} & : transition\; to\; the\; present\; cell\\
\end{array}
\right. 
\end{eqnarray}
Here, $N$ is the normalization factor.
After each pedestrian moves, procedures below are followed.  
\begin{itemize}
\item Whenever the pedestrian moves to its adjacent cell, $D$ at its origin cell is added by one.
\item $P_{ij}$ is modified according to its transition.  
\end{itemize}
\end{enumerate}

In this paper, we adopt the sequential update and reject other types of update methods (i.e. parallel and random update) for the following reasons.
Parallel update is not adopted for the simplicity of the model since it will add the step of the conflict resolution to it. Since pedestrians essentially 
avoid others in the situation discussed in this paper, we can ignore the effect of the conflict, and by doing so, we concentrate on the dynamics which comes
 from the introduction of PFF.
 When we deal with the situation in which conflicts occur (which is not discussed in this paper),
  the Proxemic Floor Field can be easily integrated into the model with the parallel update and
the conflict resolution rule as shown in \cite{evacuation}.
  Moreover, we also do not adopt the random update for the formation of inflow rule explained in the next subsection. Since the inflow speed is described 
 as the inflow probability, a change of update order leads to the inconsistency of the inflow flux.

The numbering of the updating order is assigned to each pedestrian from head to tail on the initial line as shown in Figure 4. 
This numbering corresponds to the fact that the each pedestrian mainly takes care of and is affected by other pedestrians who have already entered the area in this situation.

The transition probabilities to the neighboring cell can be replaced by
\begin{equation}
p_{ij} = N'\exp{\left\{ -(W_{ij}-W_O)-\Theta \right\}}
\end{equation}
 by modifying normalization factor $N$ without a substantial change.
This expression is explained as the following: high probabilities are allocated to the neighboring cells with relatively low $W_{ij}$, which 
lead to relatively low  $W_{ij}-W_{O}$ , i.e., gradient of total Floor Field. This formation relates the gradient of total Floor Field to the priority of
directions. The term `` $- \Theta$ " reduces moving probabilities. As explained in the previous subsection, pedestrians sometimes stop walking, ignoring the small
gradient of total Floor Field. This behavior is described by introducing this ``braking'' term; namely, the transition probability to the present cell (i.e., the probability 
with which particles stay on the present cell and do not move to neighboring cells)
 gets relatively high according to the magnitude of the threshold. Thus, transition probabilities are determined by taking $W_{ij}$ and $\Theta$ into account.

\subsection{Inflow probability}
\begin{figure}[htbp]
 \begin{center}
  \includegraphics[height=4cm,clip]{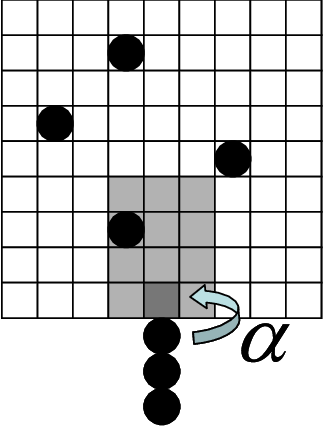}
  \hspace{15mm}
  \includegraphics[height=4cm,clip]{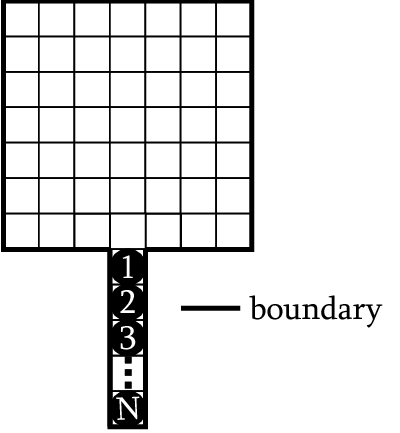}
 \end{center}
 \caption{In each time step, one pedestrian enter the area from the entrance cell indicated by the cell colored dark gray with probability $\alpha$.
 $3\times 4$ area is indicated by the area colored gray (left). Figure on the right indicates the layout of the simulation and initial position of pedestrians. The number on
 each pedestrian represents the updating order. 
 }
 \label{inf}
\end{figure}

We introduce inflow probability which describes the inflow flux of pedestrians through an entrance - given by (\ref{inflow}).
In this paper, we discuss the entrance one cell wide.
$N_{3\times 4}$ denotes the number of pedestrians in a certain area showed in Figure \ref{inf}. This expression corresponds to the fact 
that the inflow flux depends on the crowdedness of the area in front of the entrance. As as the area gets crowded, the speed of 
 pedestrian inflow shrinks. On the other hand, inflow has the maximum speed ($\alpha = 1$) if the area is free. $\rho_{cr}$ corresponds to the density where
 the inflow starts to decelerate.
\begin{eqnarray}
\alpha \left( \rho_{3\times 4} \right) &=& \min{\left\{1,\frac{1-\rho_{3\times 4}}{1-\rho_{cr}} \right\}} \label{inflow}\\
\rho_{3\times 4} &=& \frac{N_{3\times 4}}{12}
\end{eqnarray}

\section{Simulation}
By using the PFF Model, we simulate the inflow process, and investigate the characteristics of the phenomena.
In the following subsections, the number of pedestrians is fixed to 25 and we consider the square area of $L\times L$. The Static Floor Field is simply set 
as  
\begin{eqnarray}
S_{ij}&=&\left\{ 
\begin{array}{ll}
+\infty & ,(i,j):Entrance \\
0 & ,else\\
\end{array}\right.
\end{eqnarray}
to avoid the situation that a certain pedestrian stops on the entrance cell, hindering the entrance of the following pedestrian.
This assumption is natural since the pedestrians - who know the entrance of the following pedestrians, do not stop walking just after their entrance
 in the real situations.  
On the other hand, here, the Dynamic Floor Field is ignorable, since pedestrians have no reason to follow others (c.f. evacuation, counter flow, etc.), or rather avoid others.
Thus, we can set $k_D$ as $0$.
Since we are not concerned with irrational movement of pedestrians in this paper, we set enough large $k_P$ and $\Theta_{max}$.
This decreases the probabilities of transition to undesirable cells.
Therefore, the parameters remain in the model are $\Theta_{max}/k_P$ and $k_T$. We rewrite $\Theta_{max}/k_P$ as $\Theta_{max}$ in the following for 
the clarity of discussion.

\subsection{Entrance position and corner effect}\label{enPOSI}
\begin{figure}[b]
 \begin{center}
  \includegraphics[height=30mm]{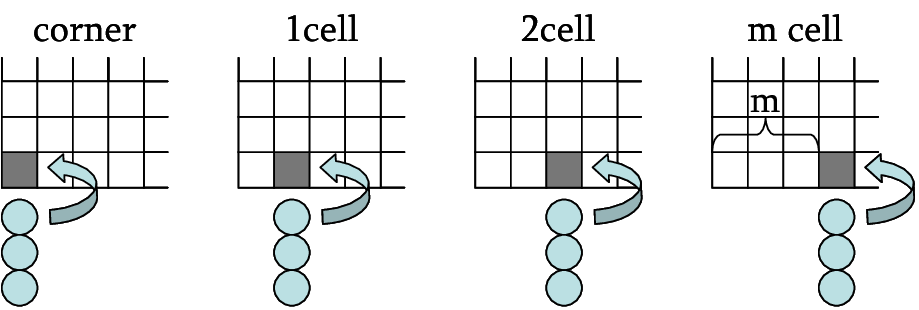}
 \end{center}
 \caption{Variety of entrance position. The entrance cell m cells away from the corner is named ``m cell" }
 \label{ent}
\end{figure}

\begin{figure}[htbp]
 \begin{center}
  \includegraphics[width=100mm]{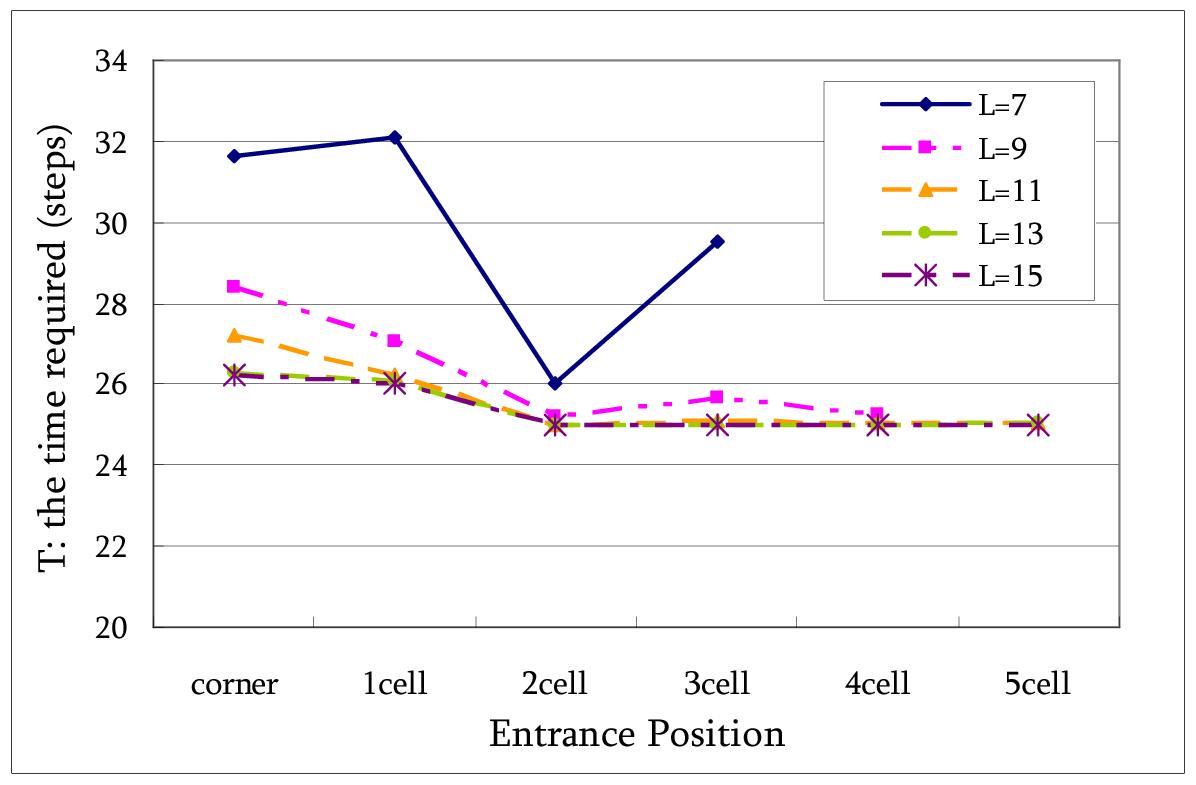}
 \end{center}
 \caption{The time required against entrance position for various $L$ in the case of $\rho_{cr}=0.2$ and $\Theta_{max}=0$.}
 \label{entsimu}
\end{figure}

Let us investigate the effect of disposition of entrance to the time required.
Here, the time required is defined to be the time step when the last pedestrian enters the room.
We set the entrance of various positions named \itshape corner/m-cell \upshape (see Figure \ref{ent}).
In order to focus on the effect of entrance disposition, we consider the condition of $\Theta_{max}=0$.
The parameter $\rho_{cr}$ is set to be 0.2 and we have acquired the simulation results as shown in Figure \ref{entsimu}.
It must be noted that only 8 cells are considered in calculating inflow probability of \itshape corner \upshape entrance.

When the area is large enough, the time required monotonically decreases as the entrance gets close to the center.
Because pedestrians tend to be attracted to boundaries for their repulsion effect, the pedestrians on the boundary hinder the inflow,
 when the entrance is near the boundary (\itshape corner\upshape).
On the other hand, when the area is small, the time required has local minimum at \itshape 2cell\upshape.
The long time required in the condition of \itshape 3cell \upshape can be explained as the following: during the inflow process, the corner cell of the area is surely occupied,
 and next cell of the corner cell (corresponds to the entrance cell of $1cell$) is not occupied for the strong repulsion force of the pedestrian occupying the corner. 
Therefore, the cell 2 cells away from the corner tends to be occupied, and hinder the inflow when the entrance is set on \itshape 3cell\upshape.

This result suggests to us the importance of entrance position especially in designing small areas for pedestrians.

\subsection{Relationship between threshold parameters and the time required}\label{para}
\begin{figure}[t]
 \begin{minipage}[t]{0.48\hsize}
  \begin{center}
   \includegraphics[width=50mm]{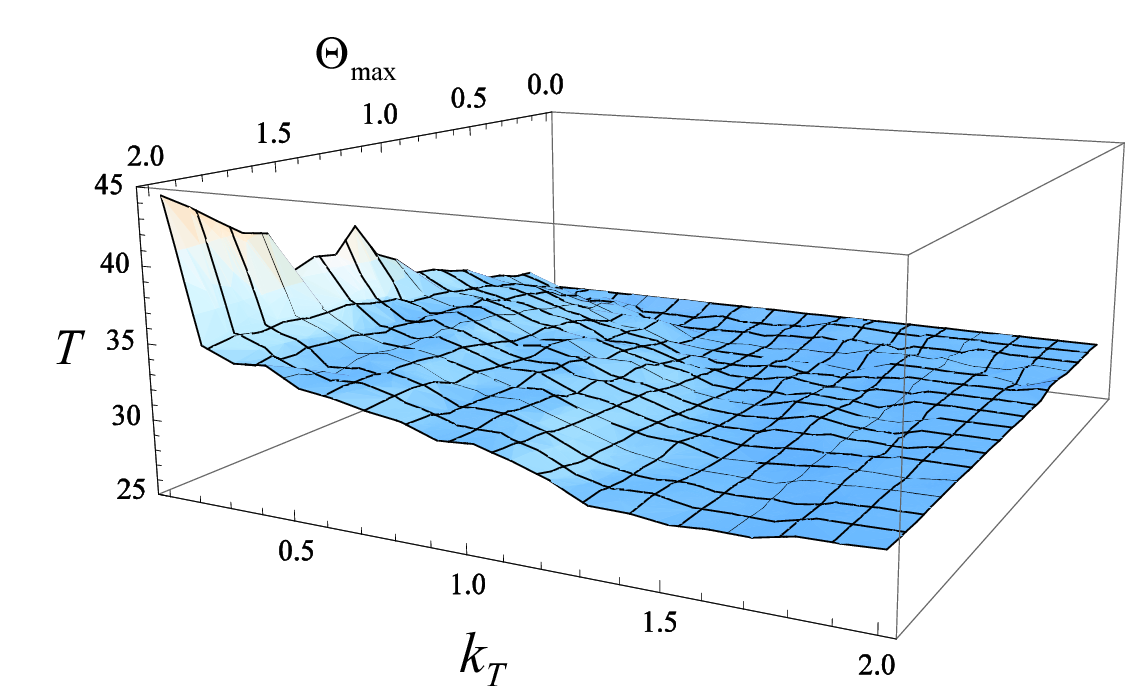}
  \end{center}
  \caption{The time required against $\Theta_{max}$ and $k_T$ in case of $L=7$, central entrance, and $\rho_{cr}=0.2$}
  \label{thres1}
 \end{minipage}
 \hfill
 \begin{minipage}[t]{0.48\hsize}
  \begin{center}
   \includegraphics[width=50mm]{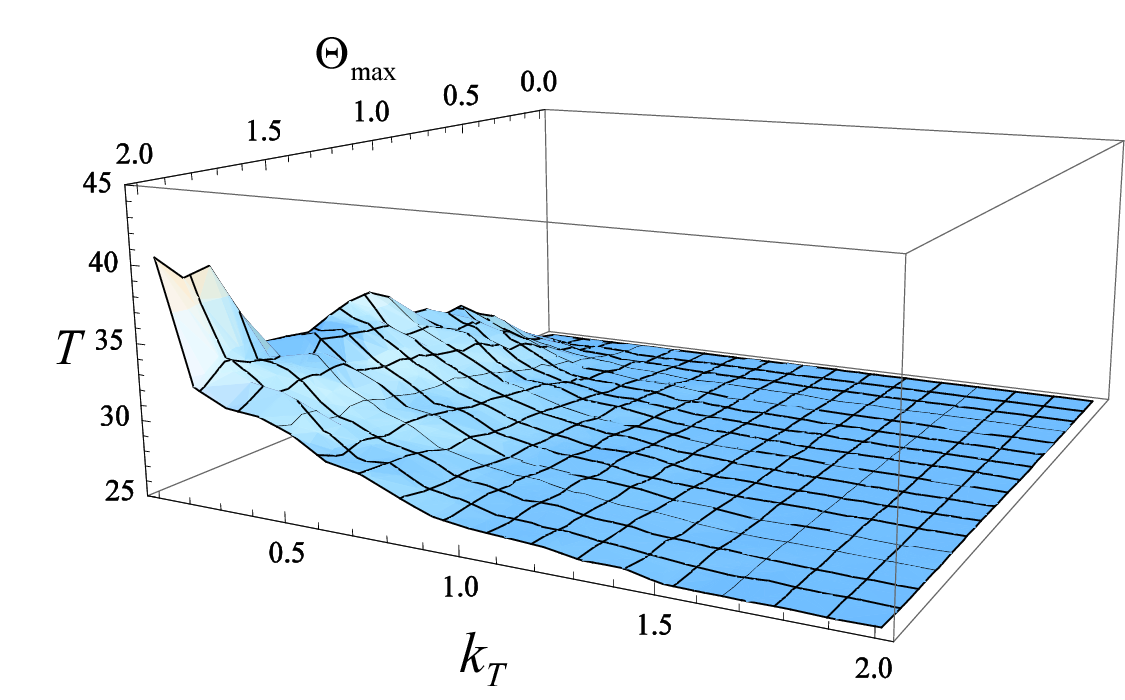}
  \end{center}
  \caption{The time required against $\Theta_{max}$ and $k_T$ in case of $L=13$, central entrance, and $\rho_{cr}=0.2$}
  \label{thres2}
 \end{minipage}
\end{figure}

The value of $\Theta_{max}$ and $k_T$ determine the shape of threshold.
Large $k_T$ gives steep decreasing threshold against $W_O$, and $\Theta_{max}$ is the coefficient of threshold.
Low $k_t$ gives gradual threshold and therefore we get relatively high values for large $W_{O}$.
Ideally rational pedestrians can be considered to ignore the gradient of $W_O$ less than or equal to threshold, thus we can describe the their sensitivity against proxemic pressure.
In the case of inflow, sensitive pedestrians (with low threshold) leads less entrance time because they do not stay just in front of entrance for their repulsion force, and the density of
$3\times 4$ area tend to be kept low. By contrast, insensitive pedestrians (with high threshold) take more time than sensitive ones because they stay in front of the entrance until the area 
gets crowded. We simulate the situation of inflow with center entrance varying $\Theta_{max}$ and $k_T$ from 0.1 to 2.0. The result of simulation is shown as Figure \ref{thres1} , \ref{thres2}.
We can see the time required decrease almost monotonically as the $\Theta_{max}$ increases or $k_t$  decreases.

\section{Analyses}

\subsection{The time required and mean field analysis}\label{analysis}
The time that it takes all the pedestrians to enter a certain area is the quantity of great concern in the context of engineering.
Here, we evaluate the time required by mean field analysis.
The inflow described in (\ref{inflow}) decelerated under the condition of congestion. In other words, when the decelerate effect is dominant, 
the area is crowded to some extent and we can then consider that the deposition of pedestrians is averaged for their repulsion forces.
On the other hand, when the area is not crowded, since pedestrians enter the area smoothly enough($\alpha=1$), mean field analysis also gives good
approximation.
By considering that the density of pedestrians is averaged and each pedestrian spreads into the area immediately, we can evaluate the time required
 where $k$ is a variable corresponding to the numbering of pedestrians.

\begin{eqnarray}
T_{ave}  &=& \sum_{k}{\frac{1}{\alpha \left(\rho_{k}\right)}} = \sum_{k}{\frac{1}{\min{\{1,\frac{1-\rho_k}{1-\rho_{cr}}\}}}} \\
\rho_{k}&=&\frac{k-1}{L^2} 
\end{eqnarray}
\begin{figure}[htbp]
 \begin{center}
  \includegraphics[width=100mm]{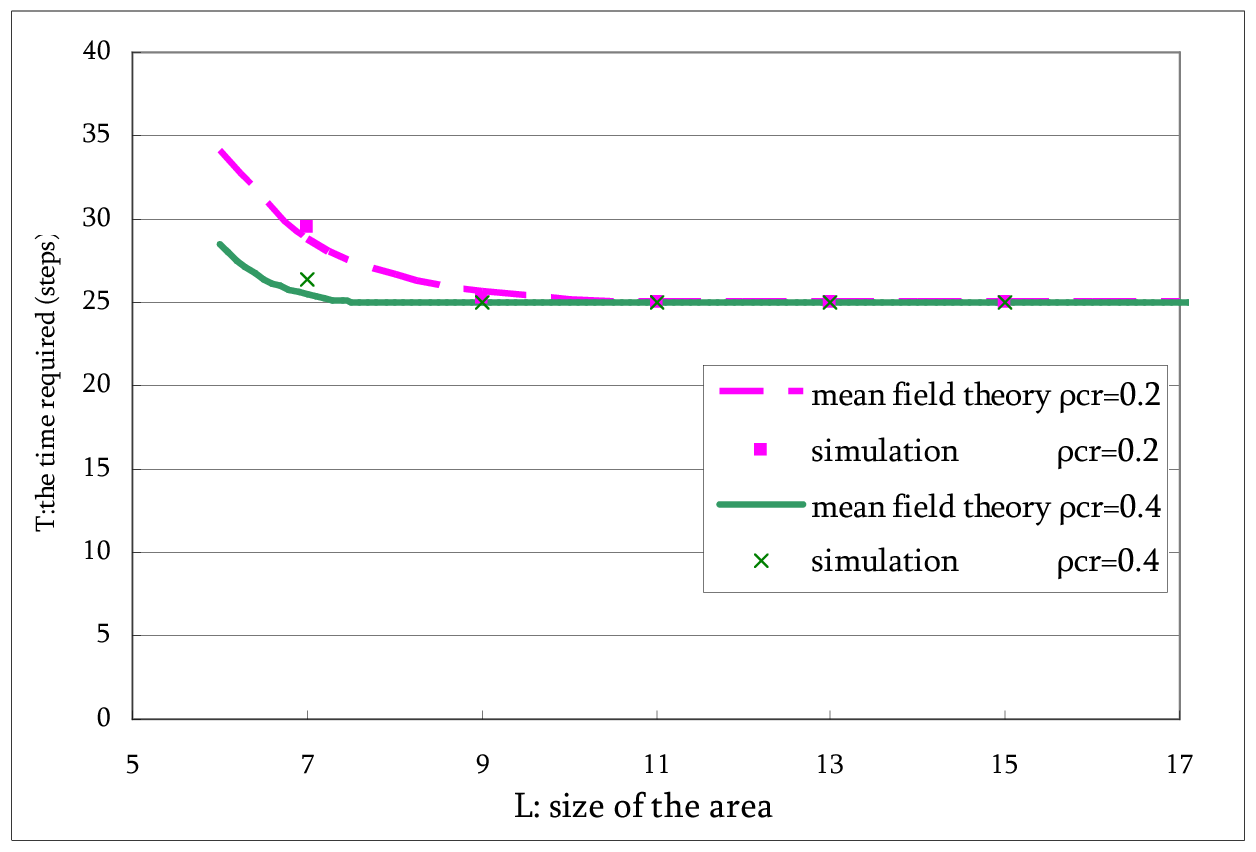}
 \end{center}
 \caption{The time required against the size of the area in case of $\rho_{cr}= 0.2, 0.4$, and $\Theta_{max}=0$}
 \label{cal}
\end{figure}
The results of simulation and approximate calculation  are shown in Figure \ref{cal}. Here, the entrance is set in the center of the boundary (see Figure \ref{inf}) and 
$\Theta_{max} = 0$ for the simplicity of the argument.
When the area is large enough, all the pedestrian enter the area in the probability $\alpha = 1$. Therefore, the time required  converges to the number of pedestrians.
By contrast, the small area leads to crowdedness and inflow is decelerated, thus the time required increases. 
In Figure  \ref{cal}, we can see that the time required for entrance increase rapidly in the cases of small area.

\subsection{Evaluation of steady state}\label{steady}
In the simulation condition we set, the steady state of the system corresponds to the situation where all the pedestrian have stopped (and no pedestrian will move once the system becomes steady),
since we prohibited irrational movement by adopting large enough $k_P$.
To consider the actual problems such as the disposition or width of the doors of train carriages, the disposition of standing passengers is highly concerned. 
Basically, uniform dispositions of pedestrians are desirable from the perspective of effective use of space.
To evaluate the desirability of disposition of pedestrians, we introduce two indices. One is \itshape spatial efficiency index \upshape $E$ defined as the following, 
\begin{equation}\label{energy}
E = \sum_{k}{W_O^k} 
\end{equation}
Where $W_O^k$ is the total Floor Field of the present cell of k-th pedestrian and the sum is taken over all the pedestrians. Since the values of total Floor Field corresponds to the uncomfortable quality of the cells, the sum of 
$W_O^k$ can be viewed as ``total proxemic stress of the system". Namely, low $E$ means the area is utilized efficiently.

The other index introduced here is \itshape unevenness \upshape$U$ defined as (\ref{uneven})
\begin{eqnarray}\label{uneven}
U & = & \sum_{l_{min}}{-p\left(l_{min}\right) \log{p\left(l_{min}\right)}} \\
p\left(l_{min}\right) & = & \# \{\, k\, |1\leq k \leq M , l^k_{min} = l_{min} \} / M
\end{eqnarray}
Here, $l^k_{min}$ is the minimum distance between $k$-th pedestrian and others, and M is the number of pedestrians.
This $U$ can be viewed as the information entropy of minimum distances, and the value of $U$ corresponds to the disorder of 
disposition of pedestrians. When the disposition is perfectly uniform (all the pedestrians give the same $l_{min}$), $U=0$.
On the other hand, when the disposition is perfectly disordered (no group of pedestrians give the same $l_{min}$),$U=\log{M}$ 

By introducing these $E$ and $U$, the characteristics of the inflow can be investigated and they can be used to categorize the inflow situations in more general cases. 
For example - as a more advanced application - the change of area shape or the introduction of obstacles can be considered.

\begin{figure}[t]
 \begin{center}
  \includegraphics[height=100mm]{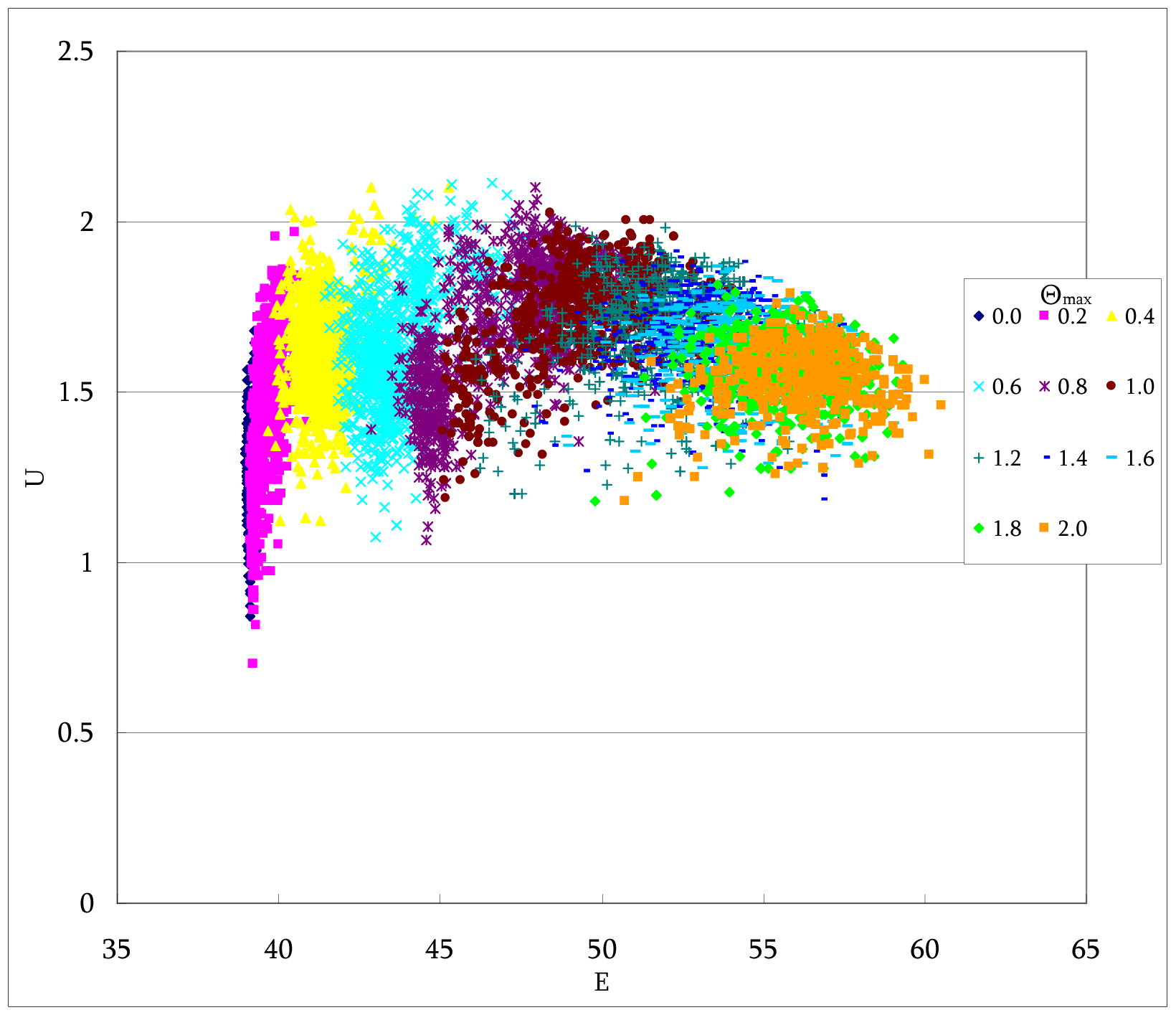}
 \end{center}
 \caption{Possible final states plotted on E-U plane for various $\Theta_{max}$ from 0.0 to 2.0 in the case of $\rho_{cr}=0.2, L=15, \alpha = 0.5$, and $k_T=1.0$}
 \label{EU}
\end{figure}

We can say that low $E$ and low $U$ give the desirable situations, and the opposite denotes the worst situations. 
Let us investigate the possible pairs of these values in the inflow situation. We plotted possible points of final states induced by many 
patterns of inflow generated by constant inflow probability $\alpha = 0.5$ (Figure \ref{EU}).
Here, we fix $k_T=1.0$ and investigate them for various $\Theta_{max}$.
$E$ increases as $\Theta_{max}$ increases large, which accounts for the situation that pedestrians stay in front of the entrance, and the space is not utilized efficiently. 
We can also see the scatter of $E$ is especially small when $\Theta_{max}$ is small. 
Turning our eyes to $U$, it should be noted that the scatter of $U$ is large when $\Theta_{max}$ is small and decreases as $\Theta_{max}$ becomes large. 
On the other hand, the average of $U$ has local maximum around $\Theta_{max} = 1.0$, and takes small value for large $\Theta_{max}$.

\section{Conclusion}
For simulating the pedestrian dynamics in which proxemic effect is dominant, we presented a extended Floor Field model 
by introducing PFF (Proxemic Floor Field) and threshold. In this paper, we focused on the dynamics of ``Inflow process" as the proxemic limit of 
pedestrian dynamics. By conducting some simulations and mean field analyses, we showed some meaningful results. In subsection \ref{enPOSI}, we  not only found that
the entrance on the corner cell causes long entrance time, but also that the local minimum of entrance time exists when the area is small. 
In subsection \ref{para}, the monotonous relationships between the entrance time, and $k_T, \Theta_{max}$ were confirmed, while in subsection \ref{analysis}, mean field analysis
roughly reconstructed the monotonous relationship between the entrance time and the area size $L$. 
Finally, subsection \ref{steady} showed the characteristics of disposition of pedestrians at steady state by introducing newly introduced indices (i.e. $energy\:  E$ and $unevenness\: U$).
Pedestrians with low threshold can lead to the final state of low $E$ and scattered $U$.
By contrast, pedestrians with high threshold can lead to high $E$ and more concentrated $U$. 

In this paper, we proposed the model which is capable of wide application and investigated the basic characteristics of it.
The problems of wider entrance, appropriate setting of $S_{ij}$ conforming to the actual situations, and the effect of area shape etc. should be investigated in the near future.

\section{Acknowledgments}
This work is financially supported by the Japan Science and Technology Agency.











\end{document}